\documentclass[aps,prc,twocolumn,showpacs,showkeys,amsmath,amssymb,superscriptaddress,nofootinbib,floatfix,longbibliography]{revtex4-2}
\usepackage{amsmath,graphicx,color,ulem}
\usepackage{epsfig,bm,slashed,hyperref}

\begin{document}

\title{Thermal and geometric normal modes of spectral fluctuations in heavy-ion collisions}

\author{Rupam Samanta}
\email{rupam.samanta@ifj.edu.pl}
\affiliation{Institute of Nuclear Physics, Polish Academy of Sciences,  31-342 Cracow, Poland}

\begin{abstract}
The transverse momentum spectrum of charged particles in ultra-relativistic heavy-ion collisions fluctuates event-by-event, encoding signatures of underlying collective dynamics. Such fluctuations originate from a combined effect of thermal and geometric fluctuations in the initial state. We present a direct decomposition of these spectral fluctuations through principal component analysis performed on the joint covariance structure of normalized spectrum, mean transverse momentum and elliptic flow squared. The first two leading modes explain 99.5\% of the total variance, and are orthogonally rotated by imposing physical constraints motivated by the initial state thermal and geometric response. The resulting thermal and geometric modes bear direct analogy with the vibrational normal modes of a linear triatomic molecule. The thermal mode entirely drives the experimentally measured $v_0(p_T)$, while the geometric mode contributes substantially to $v_{02}(p_T)$ in non-central collisions, providing a transparent explanation of its characteristic low-$p_T$ sign change. The study establishes the first physically motivated interpretation of principal component modes in the field of heavy-ion collisions and provides an experimental window into the thermo-geometric structure of the QGP initial state.   

\end{abstract}


\maketitle

\noindent {\it \bf Introduction.} Ultra-relativistic heavy-ion collision at the LHC creates a hot, dense Quark-Gluon-Plasma medium offering a unique opportunity to study the dynamics of the matter under extreme condition~\cite{Huovinen:2006jp,Ollitrault:2007du,Muller:2012zq,Heinz:2013th,Busza:2018rrf}. The characteristics of the initial fireball created immediately after the collision are accessed through the properties of the final state particles detected in the experiments. In particular, the mean transverse momentum per particle ($[p_T]$) in an event has traditionally been associated with the transverse size ($R^2$) of the fireball~\cite{Broniowski:2009fm,Bozek:2012fw,Bozek:2017elk}. However, more recent studies have established it as a reflection of the effective temperature ($T$) of the fireball characterized by its entropy density ($S/V$)~\cite{Gardim:2019xjs,Samanta:2023amp}, where $S$ is the total deposited entropy and $V$ is the effective volume. The elliptic flow ($v_2$) of the particles is governed by the shape of the initial fireball, and it is linearly related to the initial eccentricity ($\epsilon_2$) on transverse plane~\cite{Ollitrault:1992bk,Voloshin:1994mz, Sorge:1998mk,Ollitrault:2023wjk,STAR:2004jwm, ALICE:2010suc,ATLAS:2012at}.  

Event-by-event fluctuations of these final state observables are a direct consequence of initial state fluctuations. Fluctuations of $[p_T]$ are predominantly driven by the fluctuations of the effective temperature reflecting volume fluctuations at fixed entropy density~\cite{Gardim:2019xjs,Gardim:2020sma,Samanta:2023amp, Samanta:2023kfk,Alqahtani:2026edr}. Its $p_T$-differential map is constructed by correlating it with the event-by-event particle spectrum ($N(p_T)$), known as radial flow and denoted by $v_0(p_T)$~\cite{Schenke:2020uqq,Parida:2024ckk,Bhatta:2025oyp,Saha:2025nyu,Jia:2025rab,Du:2025dpu,Agarwala:2025bxs,Parida:2026kyo}. The observable $v_0(p_T)$ serves as a direct probe of the collective dynamics of the QGP fireball and has been recently measured in the experiments~\cite{ATLAS:2025ztg,ALICE:2025iud}.
 The fluctuation of event-by-event elliptic flow squared ($v_2^2$) on the other hand is primarily driven by fluctuations of event-by-event participant eccentricity squared ($\epsilon_2^2$) of the fireball~\cite{Bhalerao:2006tp,Broniowski:2007ft,Alver:2010gr,Holopainen:2010gz,Qin:2010pf,Bhalerao:2011yg,Luzum:2013yya} which is strongly correlated with the collision impact parameter. Pearson correlation coefficient between $[p_T]$ and the elliptic flow squared ($v_2^2$), called the Bo\.zek's correlator~\cite{Bozek:2016yoj}, can therefore serve as an excellent probe of the thermo-geometric coupling in the initial state  at fixed initial entropy which is equivalent to fixed multiplicity in the experiment~\cite{Bozek:2020drh,Schenke:2020uqq,Giacalone:2020byk,Giacalone:2020dln,Bozek:2021zim,Samanta:2023rbn,ALICE:2021gxt, ATLAS:2022dov,STAR:2024wgy}.  The $p_T$-differential version of this correlator is the recently proposed observable $v_{02}(p_T)$, constructed by correlating $N(p_T)$ and $v_2^2$~\cite{Parida:2025eqv}. $v_{02}(p_T)$ being a three particle correlator, is non-flow suppressed, making it a cleaner probe of the collective dynamics in comparison to $v_0(p_T)$.

The correlator $v_{02}(p_T)$ therefore serves as a differential probe of the thermo-geometric coupling in the initial state which results in correlated event-by-event fluctuations of $N(p_T)$ and $v_2^2$. As shown in reference~\cite{Parida:2025eqv}, $v_{02}$ can be decomposed into two components: one part corresponding to the fluctuations of $N(p_T)$ associated with $[p_T]$ fluctuations at fixed $v_2^2$, while the other part captures the fluctuation of spectra associated with $v_2^2$ at fixed $[p_T]$. These two components can be regarded as response coefficients or modes, and could be extracted from event-by-event or smooth hydro simulations~\cite{Parida:2025eqv}. However, the physical origin of these components and in particular characteristic double sign change of $v_{02}(p_T)$ in non-central collisions remain unexplained. Moreover, this is an indirect approach in the sense that the response coefficients can be obtained by performing a fit over events at each $p_T$-bin\footnote{This idea was suggested to the author by Jean-Yves Ollitrault}, and such decomposition is performed as a response to the final state observables, not as a response from the initial state~\footnote{It has been shown that even at fixed effective temperature, $[p_T]$ can fluctuate~\cite{Gardim:2020sma}. Such non-thermal fluctuations are small but relevant, as will be shown in this paper, in non-central collisions.} which would carry more robust physical significance. 

In this letter, we propose a direct approach to decouple the thermal and non-thermal (which we call geometric\footnote{While usual notion of geometry in the context of heavy-ion collision may indicate the fireball shape {\it i.e.} spatial anisotropy, in the present context we refer to both shape and size of the fireball by its geometry, as used throughout this article.}) modes of the event-by-event spectral fluctuations identified as a response of the corresponding initial state effects. We perform a principal component analysis~\cite{Bhalerao:2014mua,Mazeliauskas:2015efa,CMS:2017mzx} (PCA) on the joint covariance matrix constructed with normalized spectrum ($N(p_T)$), mean transverse momentum $[p_T]$ and elliptic flow squared $v_2^2$. The reason for such choice of augmented structure is to connect the principal modes obtained from PCA to the experimentally accessible observables $v_0(p_T)$ and $v_{02}(p_T)$. Without augmenting, by performing a PCA on the spectral covariance alone, the physical modes can still be approximately obtained, but no connection to the experimentally measurable observables can be established. 

Moreover, principal components are not in general physical modes. In this paper we show how one can connect the principal modes obtained from such PCA, to the physical modes of initial state response. Augmenting the spectral covariance matrix comes at the expense of a forced non-trivial mixing of the physical modes into the principal modes, since $v_{02}(p_T)$, carrying substantial geometric content, is now part of the covariance matrix. This mixing may be small in non-central collisions where the geometric source variance is naturally large, and the principal modes are already approximately aligned with the physical directions, but can be substantial in central collisions where the geometric variance is strongly suppressed, and the geometric information enters the covariance matrix predominantly through $v_{02}(p_T)$.

 We show that the original principal modes can be transformed through an orthogonal rotation by imposing physical constraints to align with the {\it thermal} and {\it non-thermal} bases. Such methods for realizing physical meaning of PC modes have been widely used and are popular in the field of meteorology~\cite{pcaRichman,pcaJolliffe}.  The thermal mode maximizes the {\it coherent} fluctuations\footnote{A coherent fluctuation of spectra refer to the changes where if number of soft particles get suppressed, hard particles production is enhanced and vice-versa. Such fluctuations are driven by pure thermal effect.} of the spectra with a single sign change or  node in $p_T$. We label the non-thermal mode as {\it geometric} since at fixed entropy density ($s(T)$), the variation of effective volume is strongly correlated with variation of total entropy resulting in eccentricity fluctuations. Such eccentricity fluctuations lead to non-trivial fluctuations of the spectra and therefore mean transverse momentum of the particles, explaining the correlation between $[p_T]$ and $v_2^2$. To gain better insights about these physical modes we draw an analogy to the {\it normal} modes of a coupled oscillator, in particular with the case of a linear triatomic molecule. Furthermore, we show that thermal and geometric modes can be experimentally accessed via the observables $v_0(p_T)$ and $v_{02}(p_T)$ by quantifying the mixing coefficients based on our analysis.       

\begin{figure}[h!]
\includegraphics[width=0.45\textwidth]{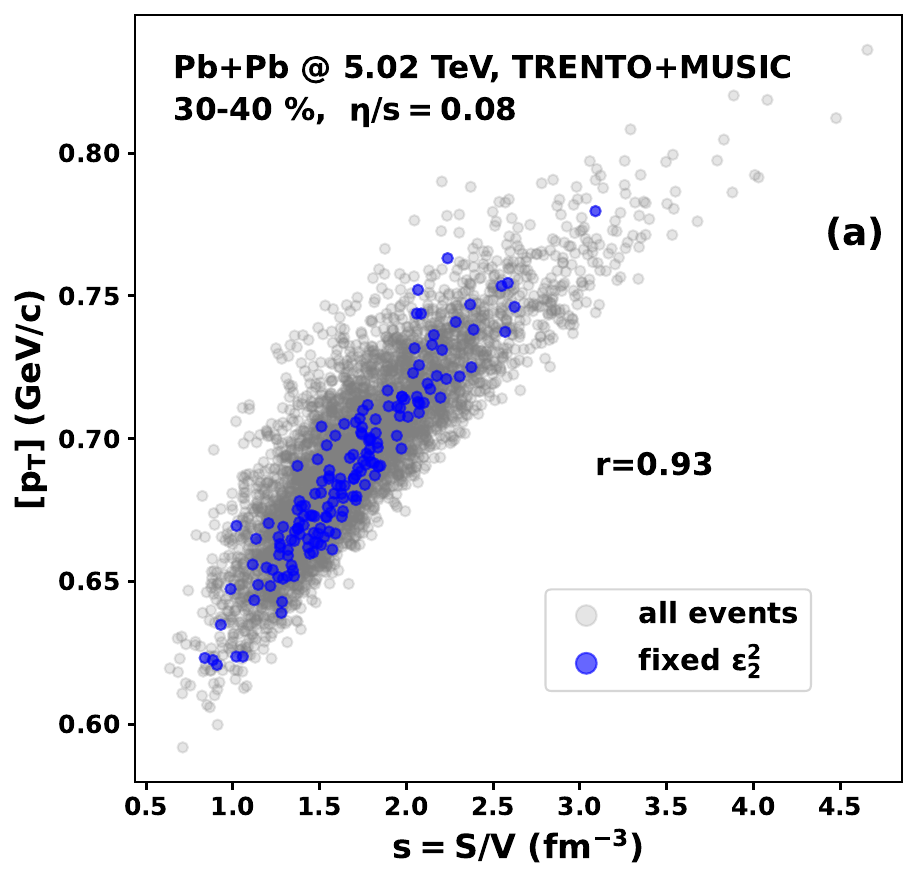} \\
\includegraphics[width=0.45\textwidth]{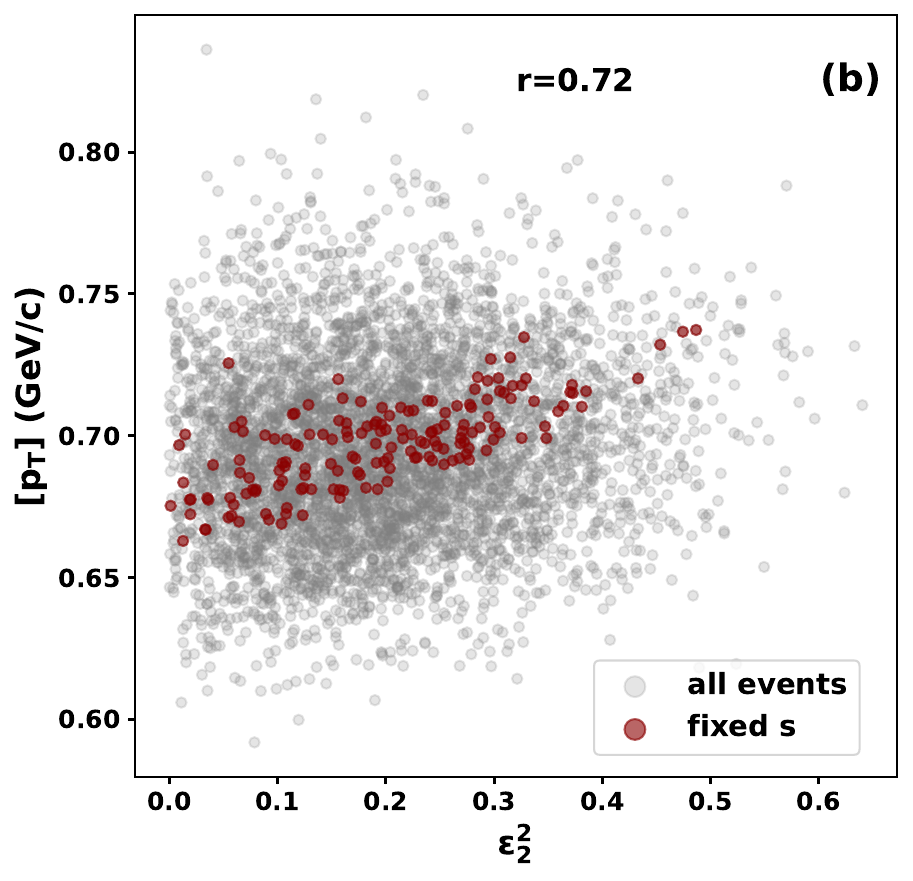}
\caption{(a) Scatter plot between event-by-event mean transverse momentum and entropy density for 5000 events in Pb+Pb collision at 5.02 TeV. Events at fixed eccentricity are shown in blue, with the full event sample shown in gray for reference. The coefficient $r$ denotes the Pearson correlation coefficient between $[p_T]$ and $s$ at fixed $\epsilon_2^2$. (b) Similar scatter plot between mean transverse momentum and eccentricity squared. The events at fixed entropy density are shown in dark red and $r$ denotes the Pearson correlation coefficient between $[p_T]$ and $\epsilon_2^2$ for those events~\label{fig:corr}}
\end{figure}

\noindent {\it \bf Spectral fluctuations and PCA.}
Traditionally principal component analysis has been used to find the dominant components in a fluctuating system. It has been applied in heavy-ion physics to analyze and identify the dominant contributions to the flow fluctuations and their correlations~\cite{Bhalerao:2014mua,Mazeliauskas:2015vea,Mazeliauskas:2015efa,Bozek:2018nne,Liu:2019jxg,Liu:2020ely,CMS:2017mzx}.
In reference~\cite{Mazeliauskas:2015efa}, the principal components of the flow fluctuations were related to the initial state predictor noting that the modes of the radial flow fluctuation can reveal crucial information on the initial state. It is a general conviction that principal components carry no physical meaning {\it a priori}--an issue already raised in~\cite{Liu:2020ely}. Therefore, a proper physical identification of principal components with initial state effects is yet to be addressed. Moreover, applying PCA on spectra alone does not establish any connection to experimentally accessible observables. In this paper we precisely address both of these issues. 

We simulate 5000 Pb+Pb collisions at a center of mass energy $\sqrt{s_{NN}}=5.02$ TeV, with fluctuating initial conditions using TRENTO~\cite{Moreland:2014oya}, and evolving them through viscous hydrodynamics using MUSIC~\cite{Schenke:2010nt} with a constant shear viscosity to entropy ratio $\eta/s=0.08$, similar to~\cite{Bozek:2021mov}. We then compute event-by-event normalized spectrum $N(p_T)\equiv(1/N)dN/dp_T$ where $N$ is the event multiplicity, mean transverse momentum per particle $[p_T]$ and elliptic flow squared $v_2^2$ following a similar method as described in~\cite{Parida:2025eqv}, to be directly used in PCA.  

We perform PCA on the normalized joint covariance matrix of $N(p_T)$, $[p_T]$ and $v_2^2$. We choose normalizations in such a way that the $N(p_T)-[p_T]$ and $N(p_T)-v_2^2$ blocks match the precise definitions of $v_0(p_T)$ and $v_{02}(p_T)$ as used in~\cite{Parida:2024ckk,Parida:2025eqv}. In particular, we normalize the $p_T$-spectrum and elliptic flow squared by their event-averages, whereas the mean transverse momentum is normalized by its standard deviation over events. With this choice, the normalized covariance matrix takes the following form:

\begin{eqnarray}
	\Sigma =
	\begin{pmatrix}
		C_{N_1,N_1} \ \dots C_{N_1,N_j} \dots  \  C_{N_1,[p_T]} \ C_{N_1,v_2^2} \\
		\vdots \\
		C_{N_i,N_1} \ \dots C_{N_i,N_j} \dots  \  C_{N_i,[p_T]} \ C_{N_i,v_2^2} \\
		\vdots \\
	\end{pmatrix}  
	\label{eq:cov_mat}
\end{eqnarray}
 
where
\begin{eqnarray}
	C_{N_i,N_j} = \frac{\langle \delta N(p_{T_i}) \ \delta N(p_{T_j})\rangle}{\langle N(p_{T_i})\rangle \langle N(p_{T_j})\rangle} \nonumber \\
	C_{N_i,[p_T]} = \frac{\langle \delta N(p_{T_i}) \delta p_T\rangle}{\langle N(p_{T_i})\rangle \sigma_{p_T}}= v_0(p_{T_i}) \nonumber \\
	\text{and} \ \ \ C_{N_i,v_2^2} = \frac{\langle \delta N(p_{T_i}) \delta v_2^2\rangle}{\langle N(p_{T_i})\rangle \langle v_2^2\rangle} = v_{02}(p_{T_i}) \ ,  
	\label{eq:mat_elem}
\end{eqnarray}
 where the indices $i,j$ denote $p_T$-bins, $\delta X = X-\langle X\rangle$ and the angular bracket $\langle \dots \rangle$ denotes average over events.
 
The principal component decomposition of $\Sigma$  can be written as:
\begin{eqnarray}
	\Sigma = E \Lambda E^T = \sum_i \lambda_i e_i e_i^T
    \label{eq:eig_decomp}
\end{eqnarray}
where $\lambda_i$'s are the eigenvalues of $\Sigma$, representing the diagonal elements of $\Lambda$, and $e_i$'s are the eigenvectors representing columns of $E$. The eigenvectors are orthonormal by construction. They have a $p_T$-dependent part and an $p_T$-independent scalar part corresponding to $[p_T]$ and $v_2^2$,
\begin{eqnarray}
	e_i=[e_i(p_{T_1}) \ e_i(p_{T_2}) \dots e_i(p_{T_j}) \dots l^{[p_T]}_i \ l^{v_2^2}_i]
\end{eqnarray}
where the $p_T$-dependent part $e_i(p_T)$ is the $i^{th}$ principal direction or {\it principal mode}, and $l_i^{[p_T]}$ and $l_i^{v_2^2}$ are the scalar components of the $i^{th}$ eigenvector corresponding to $[p_T]$ and $v_2^2$ respectively. It turns out that $99.5 \%$ variance of $\Sigma$ can be explained by the first two leading principal components.  The observables $v_0(p_T)$ and $v_{02}(p_T)$ are related to these principal components via
\begin{eqnarray}
	v_0(p_T) = \sum_i \lambda_i \ l^{[p_T]}_i \ e_i(p_T) = \sum_i \ w^{v_0}_i \  e_i (p_T) \nonumber \\
	v_{02}(p_T) = \sum_i \lambda_i \ l^{v_2^2}_i \ e_i(p_T) = \sum_i w^{v_{02}}_i \  e_i(p_T)
	\label{eq:reconst_v0v02}
\end{eqnarray}  
where $w_i = \lambda_i \ l_i$ represents the weight corresponding to the $i^{th}$ principal mode. In our particular case, only the first two principal modes ($i=1,2$) are enough for the full reconstruction of the observables in Eq.~(\ref{eq:reconst_v0v02}). 

Since the principal modes $e_i(p_T)$'s have no physical meaning {\it a priori}, one needs to impose suitable physical constraints and transform them accordingly to extract physical information by completely aligning them along pure physical modes.

\noindent {\it \bf Thermal and geometric normal modes.}
To fully decouple the principal modes into pure thermal and geometric modes, physical constraints motivated by the initial state thermodynamics must be imposed. The leading cause for the spectral fluctuation between two events is the change in effective temperature~\cite{Gardim:2019xjs,Samanta:2023amp} of the fireball which depends on the initial entropy density ($s(T)$) given by total deposited entropy ($S$) divided by the effective volume $V=(R_0^2)^{3/2}$. In a given centrality bin, at fixed eccentricity ($\epsilon_2^2$), both $S$ and $V$ can fluctuate in a weakly correlated manner. Such weak correlation suggests that at fixed $V$, $S$ can fluctuate significantly and vice versa, both of which can result in fluctuations of $s$, and therefore of $T$. Temperature fluctuations cause fluctuations of $p_T$-spectra and so of $[p_T]$. This is shown in panel (a) of Fig.~(\ref{fig:corr}), where we show the scatter plot between $[p_T]$ and $s$ at fixed $\epsilon_2^2$ in Pb+Pb collision for 30-40 \% centrality. The Pearson correlation coefficient $r$ between $[p_T]$ and $s$ for these events is very large ($r=0.93$). Such strong correlation is a manifestation of the correlation between spectra ($N(p_T)$) and effective temperature ($s(T)$). 

\begin{figure*}[h]
    \begin{center}
    \includegraphics[width=0.4\textwidth]{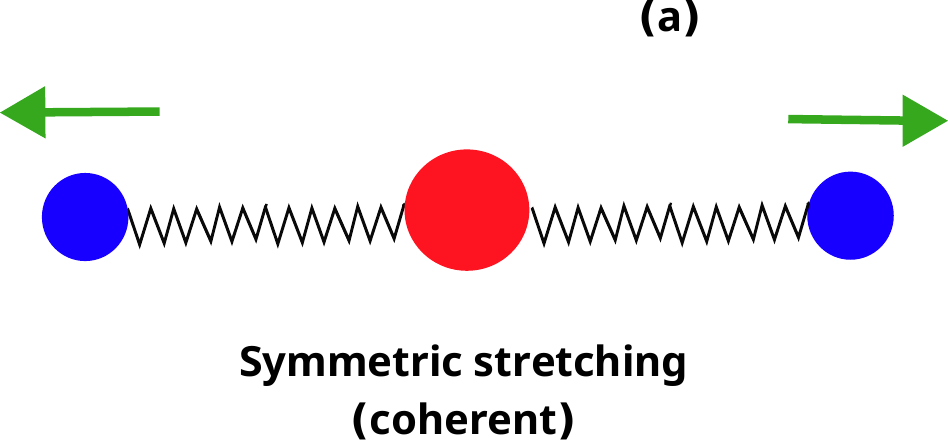}~~~~~~~~~~~~~\includegraphics[width=0.33\textwidth]{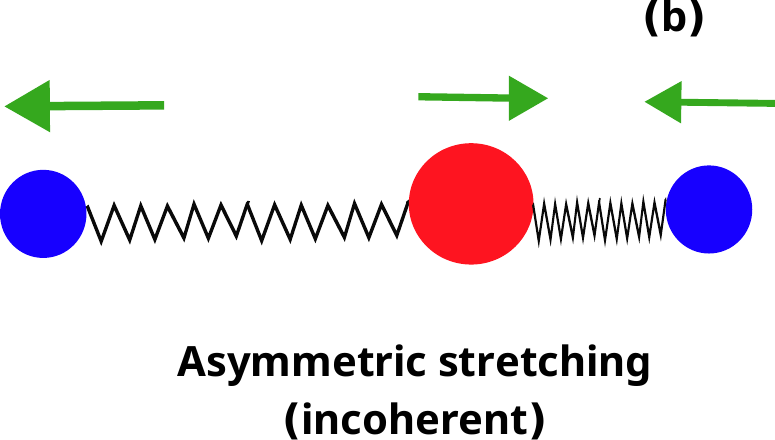}
    \vskip 1 cm
    \includegraphics[width=0.38\textwidth]{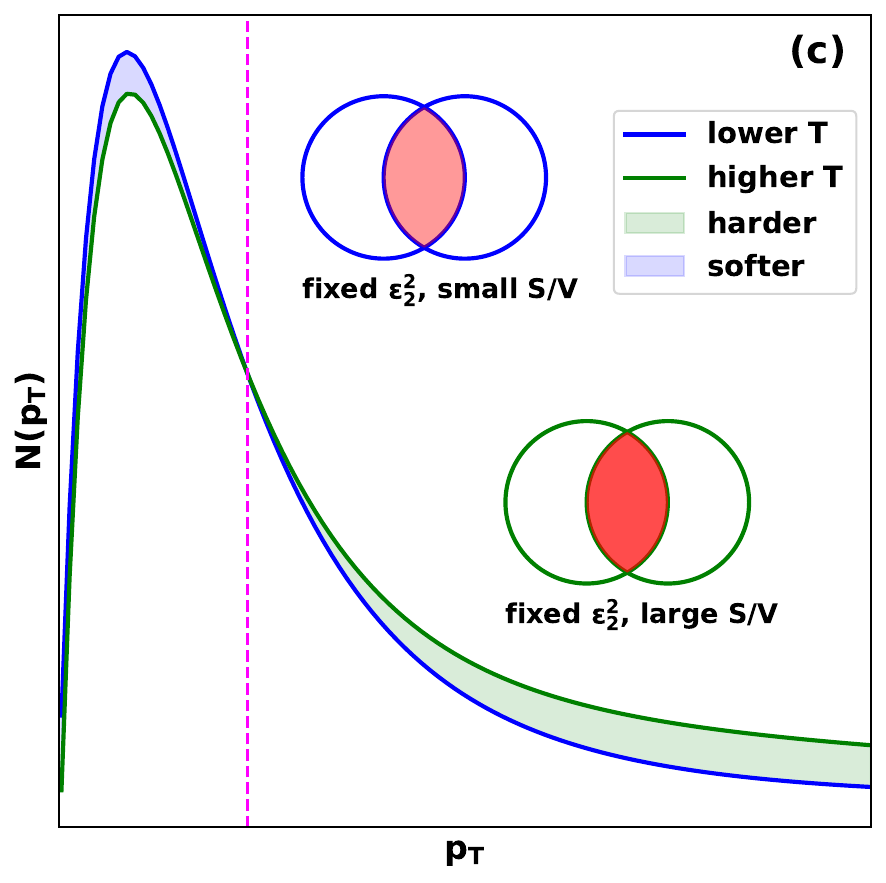}~~~~~~~~~~~~\includegraphics[width=0.38\textwidth]{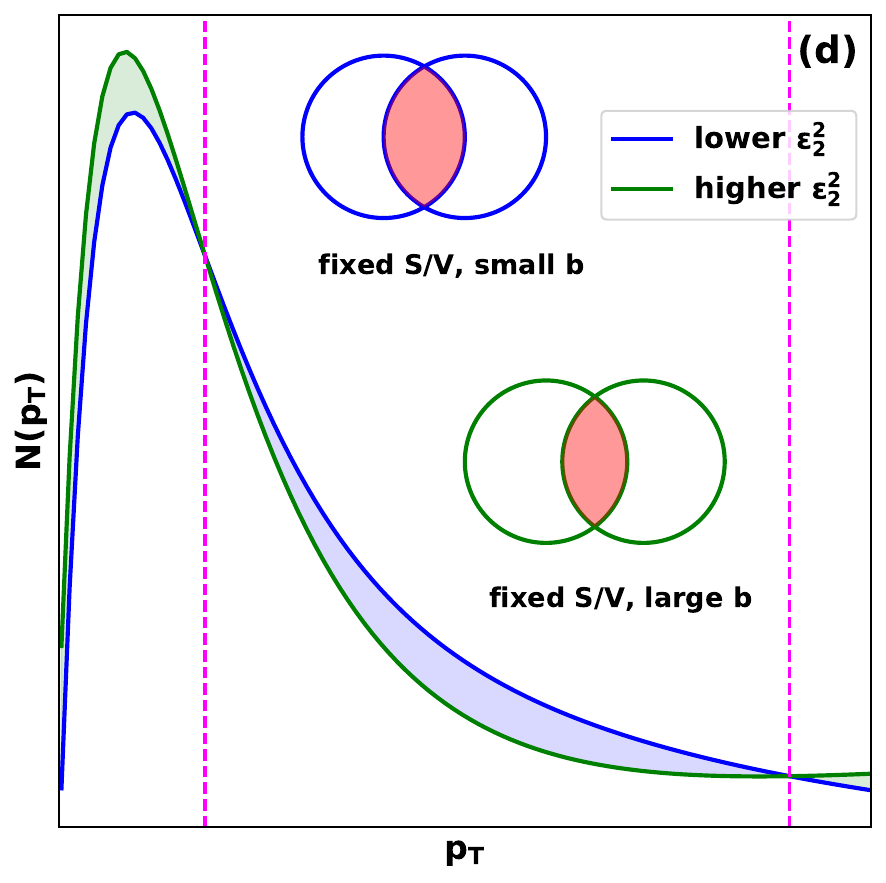}
    \vskip 1 cm
    \includegraphics[width=0.42\textwidth]{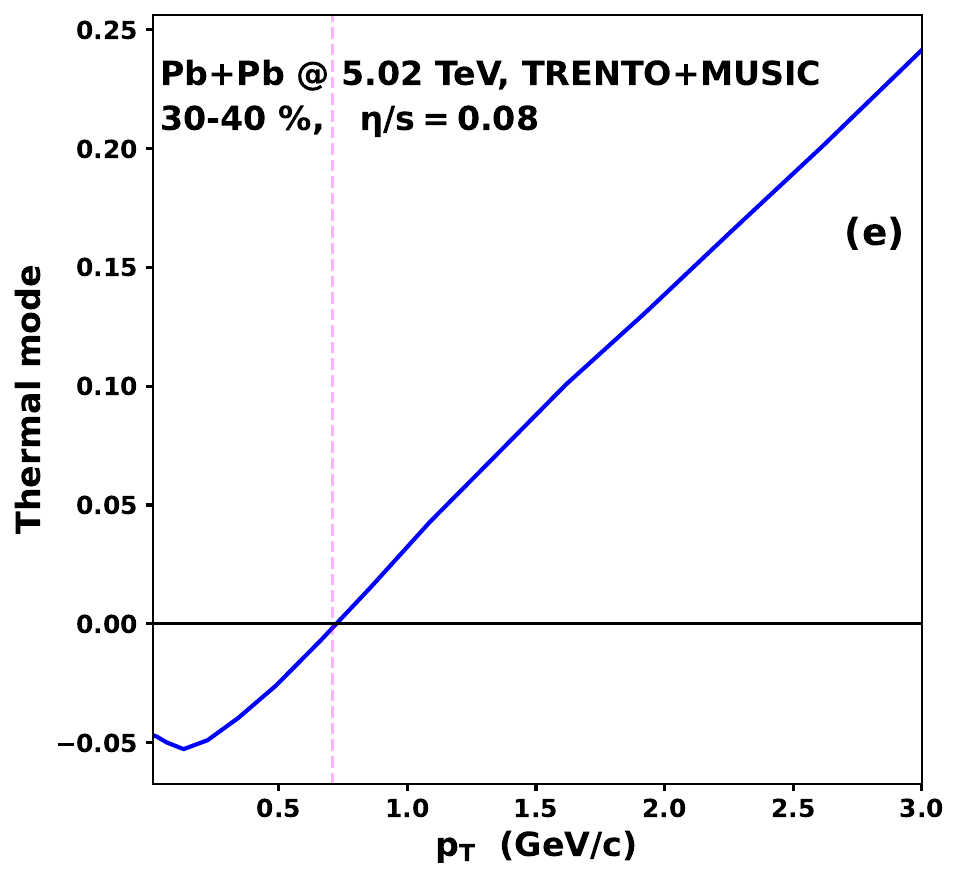}~~~~~~~\includegraphics[width=0.42\textwidth]{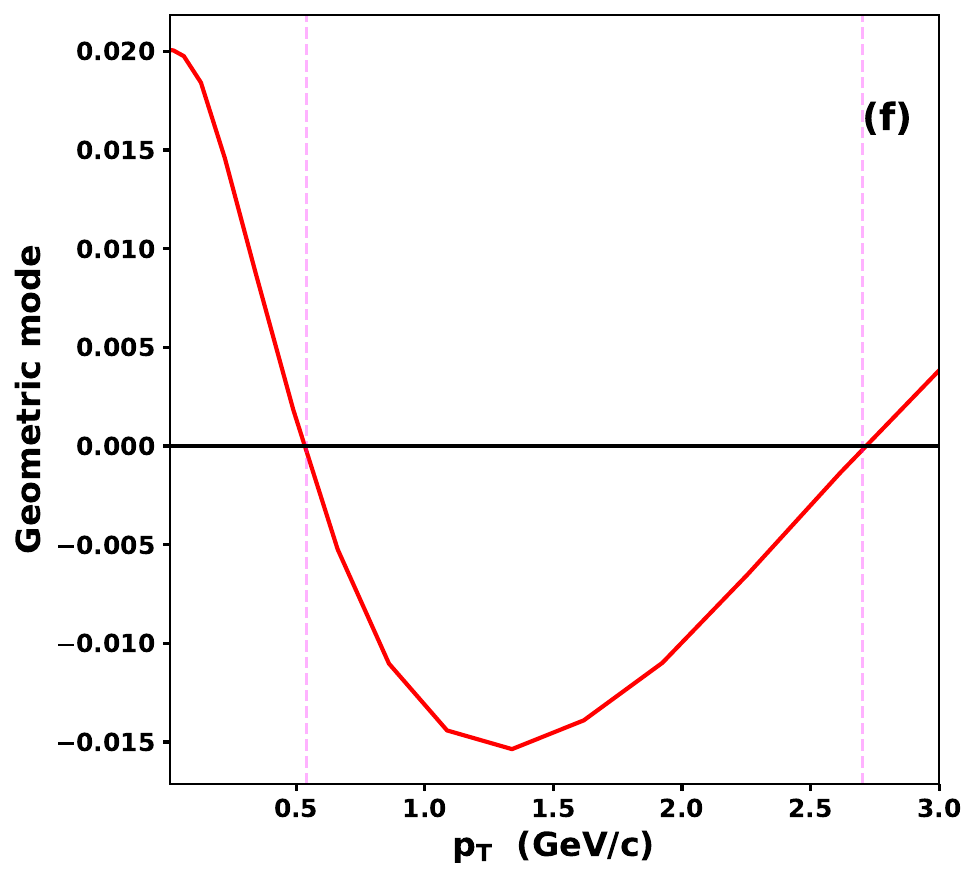}
    \caption{(a) Cartoon of the symmetric stretching (coherent) mode of vibration in a linear triatomic molecule, where the middle atom denoted by red remains at rest while the atoms on wither side, denoted by blue, move in opposite directions. (b) Asymmetric vibrational mode of the molecule where the atoms on either side move in same direction and the middle atom moves in opposite direction. (c) Pictorial depiction of fluctuation of normalized transverse momentum spectra in two events caused by thermal fluctuations due to entropy density fluctuations at fixed eccentricity. An increase in effective temperature in the initial state leads to smaller number of soft particles and larger number of hard particles, a change that is coherent in nature pivoted at $\langle p_T \rangle$ denoted by the vertical dashed line. The hotter fireball denoted by bright red color correspond to the green spectrum, whereas the colder fireball is represented by light red color corresponding to the spectrum in blue. (d) Similar cartoon of spectra and initial fireball at fixed $S/V$, causing the spectra to intersect two times. To maintain same effective temperature the total entropy and effective volume needs to change simultaneously. A decrease in effective volume is an outcome of larger impact parameter, hence larger $\epsilon_2^2$. (e) Thermal normal mode of spectral fluctuation in Pb+Pb collision at 5.02 TeV for 30-40 \% centrality, plotted as a function of $p_T$ having a single node. (f) Geometric normal mode having characteristic double sign change in same centrality class~\label{fig:modes}}
    \end{center}
\end{figure*}

An increase in temperature always leads to production of less number of soft particles and more number of hard particles~\cite{Parida:2024ckk,ATLAS:2025ztg}--hereby referred as coherent change of the spectrum. Such changes are pivoted at $\langle [p_T] \rangle \equiv \langle p_T \rangle$. This is illustrated by the cartoon in panel (c) of Fig.~(\ref{fig:modes}). We identify such mode of the spectral fluctuations as the {\it coherent mode} having a single node in $p_T$ in an analogy to symmetric stretching of a linear triatomic molecule (e.g. $CO_2$), illustrated by the cartoon in panel (a) of Fig.~(\ref{fig:modes}). As in the symmetric stretching mode, the atoms in the either sides of the triatomic molecule moves in opposite direction keeping the middle one fixed, in the coherent mode of spectral fluctuation with the increase of effective temperature of the fireball the normalized spectra below $\langle p_T \rangle$ moves downwards and the part above it moves upwards.

To obtain such a mode from our original principal modes ($ e_i (p_T)$), we apply an orthogonal rotation on the original modes such that the event-by-event coherent changes in the spectra is maximized, hence called the {\it maximal coherence criterion}. In particular, we make the following transformation on the original principal component bases for each $p_T$ bin: 
\begin{eqnarray}
	\begin{pmatrix}
	   e_T  \\
	   e_G  \\
	\end{pmatrix} = 
    \begin{pmatrix}
    	\cos \theta & \sin \theta  \\
    	-\sin \theta & \cos \theta  \\
    \end{pmatrix} \ 
    \begin{pmatrix}
    	 e_1  \\
    	 e_2  \\
    \end{pmatrix}
    \label{eq:rotation}
\end{eqnarray}
where the $p_T$ dependence of the modes on either side of the equation is implicit. The angle $\theta$ is the global optimal angle of rotation applied uniformly across all $p_T$-bins and is obtained by a weighted least squares of the optimal angles corresponding to each $p_T$-bin. The mathematical justification for such rotation, rooted in rigorous connection between PC eigenmodes and the initial state thermal and geometric response functions (physical modes), is provided in Appendix~\ref{s:relation}. The global optimal angle $\theta$ is derived in Appendix~\ref{s:rotation} via the maximal coherence criterion, and independently determined in Appendix~\ref{s:rotation_alt} by imposing a physical condition. The two methods agree to within 3 \% of the total rotation range. 

By maximizing the coherent deformation of the spectrum we contain all the fluctuations coming out of thermal effect in a single {\it pure mode} $ e_T$, hence labeled as {\it thermal mode}. The thermal mode is shown in panel (e) of Fig.~\ref{fig:modes}. We show results for 30-40 \% centrality where the original principal modes are apprioximately aligned along physical directions, hence the orthogonal rotation plays small role \footnote{It should be noted that such orthogonal rotation is a general requirement according to the rigorous relationship between final and initial state fluctuations shown in Appendix~\ref{s:relation}, irrespective of the mixing between the physical modes into principal modes. Such rotation is expected to play most consequential role in central collisions where the mixing is substantial.} although essential for the physical identification. It is negative below $\langle p_T \rangle$ signifying the negative correlation of soft particles with increase in effective temperature, and positive above  $\langle p_T \rangle$.

The second rotated mode $e_G$ represents all residual spectral fluctuations that do not have a thermal origin. It is shown in panel (f) of  Fig.~\ref{fig:modes}. This mode has double sign-change, or two nodes. The first one occurs slightly below $\langle p_T \rangle$ and the second one occurs at high $p_T$ beyond 2.5 GeV. The mode implies a deformation of the spectra where the soft particles are positively correlated with the changes in the initial state, intermediate particles are negatively correlated, and hard particles are again positively correlated. Physically, at fixed temperature, a fireball with larger eccentricity develops stronger pressure gradients along its minor axis, resulting in a modified radial-flow profile and a small change in $[p_T]$. Such geometry-induced spectral redistribution among different $p_T$ regions produces the characteristic two-nodal structure. This is analogous to the asymmetric stretching of a linear triatomic molecule, where the atoms on either side move in the same direction, but the middle atom moves in opposite direction, resulting in an incoherent mode of vibration. This is illustrated by the cartoon in panel (b) of Fig.~\ref{fig:modes}.

At fixed effective temperature, a proportionate (perfect correlation) change in $S$ and $V$ owing to impact parameter variation leads to fluctuations of $\epsilon_2^2$. The effect is shown in panel (b) of Fig.~\ref{fig:corr} presenting a scatter plot between $[p_T]$ and $\epsilon_2^2$ at fixed $s$ implying fixed effective temperature. There exist a strong correlation ($r=0.72$) between the two in 30-40 \% centrality. The corresponding variation of $[p_T]$ is smaller as compared to thermal fluctuations, but sufficiently enough to enforce a non-trivial change in $p_T$-spectra. The scenario is illustrated by the cartoon in panel (d) of Fig.~\ref{fig:modes}, where two spectra corresponding to two events cross at two different $p_T$. Therefore, two events can have same $S/V$, hence same $T$ but different $\epsilon_2^2$ due to different $b$, hence different geometry. We label this second rotated mode $e_G$ as the {\it geometric mode} of spectral fluctuation, as it originates from the changes in shape and size of the fireball. 

\noindent {\it \bf Relevance to heavy-ion experiments.} It is straightforward to reconstruct the observables $v_0(p_T)$ and $v_{02}(p_T)$ in the rotated bases as,
\begin{eqnarray}
	v_0(p_T)  = \sum_i \ \tilde{w}^{v_0}_i \  \tilde{e}_i(p_T) \nonumber \\
	v_{02}(p_T) =  \sum_i \tilde{w}^{v_{02}}_i \ \tilde{e}_i(p_T)
	\label{eq:reconst_v0v02_rotated}
\end{eqnarray}
where $\tilde{e}_1\equiv e_T$, $\tilde{e}_2\equiv e_G$ and $\tilde{w}_i$'s are the weights corresponding to the thermal and geometric normal modes. They are obtained by applying the same orthogonal rotation as in Eq.~(\ref{eq:rotation}) on the weights corresponding to the original principal modes in Eq.~(\ref{eq:reconst_v0v02}).
\begin{figure}[h!]
\includegraphics[width=0.45\textwidth]{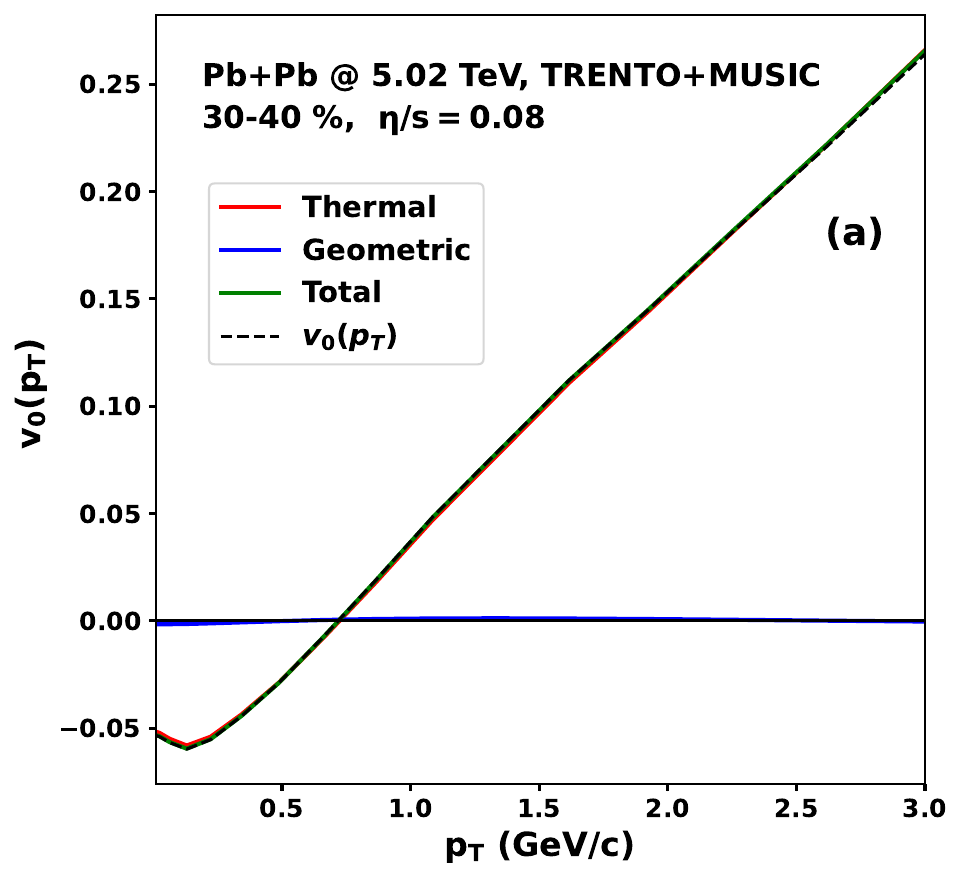} \\
\includegraphics[width=0.45\textwidth]{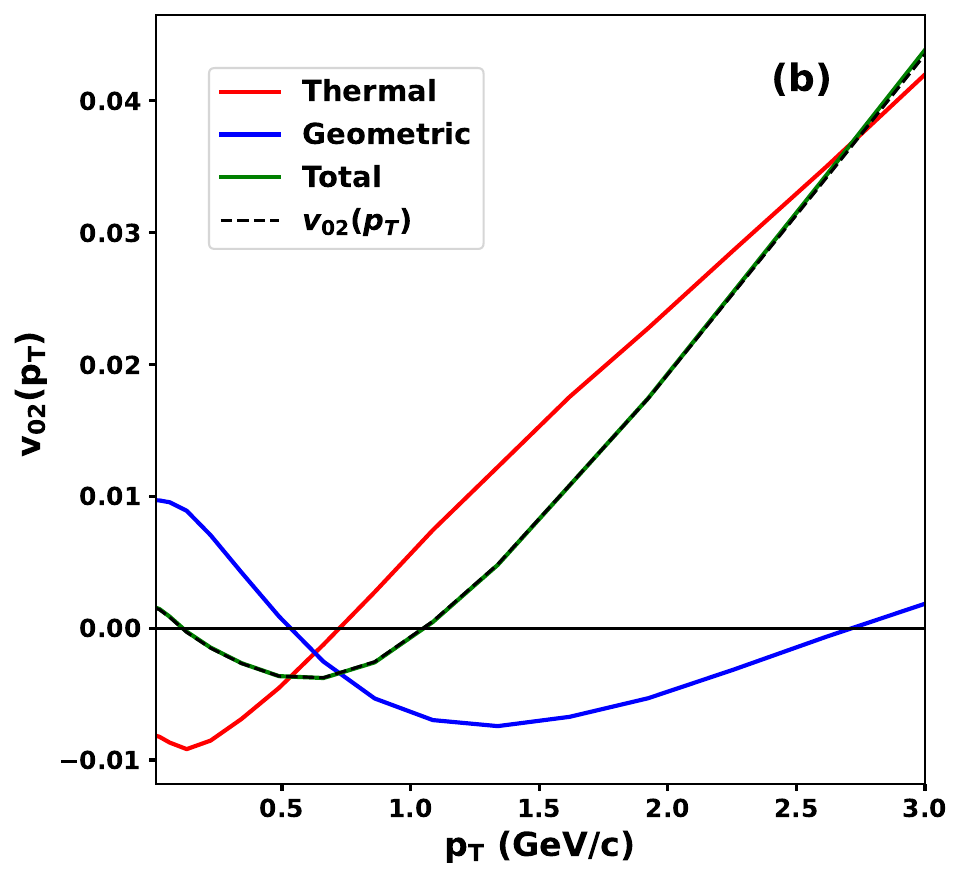}
\caption{(a) Results for the thermal (red) and geometric (blue) contributions to $v_0 (p_T)$ in Pb+Pb collision at 5.02 TeV in 30-40\% centrality class, as obtained in Eq.~(\ref{eq:reconst_v0v02_rotated}). The green curve represents the sum of the two. The black dashed line represents the result computed directly using the original definition as given in Eq.~(\ref{eq:mat_elem}). (b) Thermal (red) and geometric (blue) components of $v_{02}(p_T)$ along with their sum (green). The black dashed line represents the result computed directly using the definition in Eq.~(\ref{eq:mat_elem}). ~\label{fig:v0v02}}
\end{figure}

In Fig.~\ref{fig:v0v02}, we present the results for $v_0(p_T)$ and $v_{02}(p_T)$ reconstructed using Eq.~(\ref{eq:reconst_v0v02_rotated}), along with their thermal and geometric counterparts. We also present the results for these observables directly computed from hydro results, to establish the fact that they can be fully reconstructed by using the first two leading modes. We show results for 30-40 \% centrality, for which the geometric component has substantial effect on $v_{02}(p_T)$ leading to its sign change at low $p_T$. The low-$p_T$ sign-change may not be present in central collision due to the saturation of impact parameter fluctuations towards central collisions~\cite{Samanta:2023amp,Samanta:2023kfk}, resulting in a smaller contribution from the geometric mode. A detailed study of the centrality dependence of the modes is left for a future study. 

It can be seen from panel (a) of Fig.~\ref{fig:v0v02} that $v_0(p_T)$ is entirely driven by the thermal mode, with almost no contribution from the geometric mode. This is because the corresponding weight $\tilde{w}_2^{v_0} \equiv \tilde{w}_G^{v_0}$ constitutes just $\approx$ 7\% of the total and the geometric mode $e_{G}$ is itself suppressed by an order of magnitude c.f. Fig.~\ref{fig:modes}(f). This is also consistent with the results at fixed $b$~\cite{Parida:2024ckk}, where thermal fluctuations dominate. It follows that within a given centrality class, $v_0(p_T)$ directly maps the thermal normal mode of the event-by-event spectra and their relation can be written as
\begin{eqnarray}
    e_T(p_T) \approx \alpha \  v_0(p_T) \ .
    \label{eq:v0_thermal}
\end{eqnarray}
where $\alpha$ is the normalization coefficient reflecting the magnitude of thermal mode. This fact is used for the estimation of $\theta$ in Appendix~\ref{s:rotation_alt}. From our analysis, $\alpha \approx 0.9$ for the 30-40 \% centrality bin. Therefore, the measured $v_0(p_T)$ ~\cite{ATLAS:2025ztg,ALICE:2025iud}, provides a robust experimental proxy of the underlying thermal mode of spectral fluctuations. 

Panel (b) of Fig.~\ref{fig:v0v02} shows $v_{02}(p_T)$ along with its thermal and geometric components. It is clearly visible that the geometric component has a substantial contribution, forcing a double sign change in $v_{02}(p_T)$, particularly at low $p_T$. Quantitatively, the weight corresponding to the geometric component carries 75\% of the total weight, resulting in an enhanced contribution from geometric mode $e_G$ in non-central collisions. This is an artifact of the strong correlation shown in panel (b) of Fig.~\ref{fig:corr}. Additionally in both panels of Fig.~\ref{fig:v0v02}, directly computed results for $v_0(p_T)$ and $v_{02}(p_T)$ are shown to establish the fact that only first two leading modes are enough for the full reconstruction of the obervables.  Unlike the thermal mode, the geometric mode cannot be directly realized from experimental measurement of $v_{02}(p_T)$ due to substantial mixing of modes. However, by measuring $v_{02}(p_T)$ as guided in ~\cite{Parida:2025eqv}, it can be indirectly extracted via the following relation:
\begin{eqnarray}
	 e_G (p_T) = \beta_1 v_{02}(p_T)- \beta_2 v_0(p_T)
\end{eqnarray}
where the mixing coefficients $\beta_1 \approx 2 $ and $\beta_2 \approx 0.3 $, as obtained from our analysis for the 30-40\% centrality class. 

Since the above described method is an indirect way of extracting the thermal and geometric modes of spectral fluctuations subject to some model inputs, a potential direct way of accessing these modes would involve repeating the same PCA on experimental {\it data covariance matrix} of normalized spectrum, $[p_T]$ and $v_2^2$, following a similar procedure as outlined in reference~\cite{CMS:2017mzx}. For this direct measurement one needs to design four well-separated pseudorapidity windows for four group of sub-events to construct the required data covariance matrix~\cite{Parida:2025eqv}. More details of such a experimental procedure are provided in Appendix~\ref{s:expt_method}.  

\noindent {\it \bf Conclusions.} In this letter we have established that event-by-event fluctuations of transverse momentum spectra in heavy-ion collisions decompose into two physically distinct normal modes: thermal and geometric, in direct analogy to the vibrational normal modes of a linear triatomic molecule. The thermal mode captures coherent spectral deformation driven by entropy density fluctuations at fixed eccentricity, while the geometric mode represents the incoherent residual deformation arising from eccentricity and size fluctuations at fixed temperature. Both of these effects are supported by direct empirical evidence from event-by-event correlations between final and initial state. Such decomposition can be achieved by performing principal component analysis on the augmented joint covariance matrix of the $p_T$-spectra, mean transverse momentum and elliptic flow squared, which establishes direct connection to the experimentally accessible observables $v_0(p_T)$ and $v_{02}(p_T)$. The thermal and geometric modes are then obtained via an orthogonal rotation of the original principal modes, a method that is mathematically grounded in the rigorous connection between PC eigenmodes and initial state thermal and geometric response functions. 

The observable $v_0(p_T)$ maps almost entirely the thermal mode and $v_{02}(p_T)$ receives substantial contribution from geometric mode in non-central collisions, directly explaining its characteristic low-$p_T$ sign change. This study further highlights the importance of measuring $v_{02}(p_T)$ in experiment in immediate future. While the physical modes can be indirectly realized by measuring  $v_0(p_T)$ and $v_{02}(p_T)$ up to some model inputs, a direct experimental extraction of these modes is also possible following a data-driven PCA method. Beyond the specific results presented here, the maximal coherence rotation criterion is conceptually general and can be extended to other collective probes, reframing the spectral fluctuation problem: rather than asking what $v_0(p_T)$ and $v_{02}(p_T)$ measure independently, the natural question becomes how the thermal and geometric normal modes project onto experimental observables, opening a direct and systematic window into thermo-geomteric structure of QGP.

\noindent {\it \bf Acknowledgements.} The author thanks Jean-Yves Ollitrault for stimulating physics discussion and constructive feedback that significantly improved the manuscript, and Tribhuban Parida for engaging discussions and feedback on the draft. He also thanks Wojciech Broniowski for feedback and discussions on the manuscript, and especially thanks Piotr Bo\.zek for discussions on fluctuations of $p_T$-spectra and experimental method for extraction of the PC modes. The author acknowledges support from the Polish National Science Center grant 2023/51/B/ST2/01625, and thanks IPhT Saclay for hospitality during a research visit where a significant part of this work was carried out.

\bibliography{ref}

\appendix

\section{Relation between final and initial state}
\label{s:relation}
In our analysis the event-by-event fluctuation of final state observables can be written in matrix form:
\begin{eqnarray}
    \delta X =  \left( \delta N(p_{T_1}) \ \dots \ \delta N(p_{T_i})  \dots  \ \delta p_T \ \delta v_{2}^2 \right)
\end{eqnarray}
where $i$ runs from 1 to $m$ with $m$ being the number of $p_T$-bins and dimension of $X$ is $m+2$. One can relate this to initial state fluctuation as,

\begin{eqnarray}
   \begin{pmatrix}
        \delta N(p_{T_1}) \\ 
        \vdots \\ 
        \delta N(p_{T_i}) \\  
        \vdots  \\ 
        \delta p_T \\ 
        \delta v_{2}^2
   \end{pmatrix} =
   \begin{pmatrix}
         \alpha_1(p_{T_1}) &  \alpha_2(p_{T_1}) \\ 
         \vdots \\ 
         \alpha_1 (p_{T_i}) & \alpha_2 (p_{T_i})\\  
         \vdots  \\ 
         \beta_1 & \beta_2 \\ 
         \gamma_1 & \gamma_2 \\
   \end{pmatrix} 
   \begin{pmatrix}
         \delta T \\
         \delta \epsilon_2^2
   \end{pmatrix}
\end{eqnarray}
which can be written as,
\begin{eqnarray}
    \delta X = R \ \delta S \ .
\end{eqnarray}
where $R$ is the response matrix with its columns representing the thermal and geometric response modes, and $\delta S$ represents the corresponding thermal and eccentricity fluctuations at initial state. Then for PCA one compute the covariance matrix,
\begin{eqnarray}
    \Sigma = \frac{1}{n-1} \sum_i \delta X_i \ \delta X_i^T = R \ \Sigma_S R^T
\end{eqnarray}
where the sum is over the number of events and resulting covariance matrix $\Sigma$ having the form as shown in Eq.~\ref{eq:cov_mat}. $\Sigma_s$ is the source covariance matrix and can be written as, $\Sigma_S=\frac{1}{n-1} \sum_i \delta S_i \ \delta S_i^T$.  Then the eigen-decomposition in Eq.~(\ref{eq:eig_decomp}) can be related to the initial state as, 
\begin{eqnarray}
    E \Lambda E^T = R \ \Sigma_S R^T \ .
\end{eqnarray}
Since the matrix $E$ is orthonormal, the above equation can be written as, 
 \begin{eqnarray}
     \Lambda = (E^TR) \ \Sigma_S (R^T E) = O^T \Sigma_S O
\end{eqnarray}
since $\Lambda$ is diagonal, the matrix $R^TE \equiv O $ must be orthonormal which diagonalizes $\Sigma_S$. So the columns (the bases) of $R$ and $E$ are related by orthogonal transformation. In other words, since the columns of $E$ and $R$ spans the same subspace and these columns are orthogonal by construction, they must be related by some orthogonal transformation. Therefore, one writes 
\begin{eqnarray}
&R^T & = O E^T \\
\\
\text{or,} \ \ 
    &\begin{pmatrix}
	   \alpha_1  \\
	   \alpha_2  \\
	\end{pmatrix} &= 
     \begin{pmatrix}
    	\cos \theta & \sin \theta  \\
    	-\sin \theta & \cos \theta  \\
    \end{pmatrix}
    \begin{pmatrix}
    	 e_1  \\
    	 e_2  \\
    \end{pmatrix}    
\end{eqnarray}
which is exactly being done above where $\alpha_1 \equiv e_T$ and $\alpha_2 \equiv e_G$.

\section{Optimal angle for orthogonal rotation}
\label{s:rotation}

The orthogonal rotation transforms the principal modes into physical thermal and geometric modes. For the thermal mode,
\begin{eqnarray}
	e_T (p_{T})= \cos \theta \ e_1(p_{T}) + \sin \theta \ e_2(p_{T})
\end{eqnarray}
where $\theta$ is a single global rotation angle  applied uniformly across all $p_T$ bins. A global angle is required by construction: since $e_T$ and $e_G$ must be fixed linear combinations of $e_1$ and $e_2$ independent of $p_T$. A $p_T$-dependent rotation angle would yield $p_T$-dependent basis vectors, destroying the normal mode interpretation.

The global angle $\theta$ is determined by the maximal coherence criterion which imposes the condition that $e_T$ contains all thermal fluctuations within a single mode. Since pure thermal fluctuations produce coherent single-node spectral deformation pivoted at $\langle p_T \rangle$, this is equivalent to maximizing $e_T(p_{T_i})$ at each bin independently:
\begin{eqnarray}
    \frac{\partial e_T(p_{T_i})}{\partial \theta_i} =0
\end{eqnarray}
which returns the bin-optimal angle
\begin{eqnarray}
    \theta_i = \tan^{-1} \left( \frac{e_2(p_{T_i})}{e_1(p_{T_i})} \right) \ .
\end{eqnarray}

Since different $p_T$ bins yield different $\theta_i$, we seek a single global $\theta$ that best approximates all bin-specific angles simultaneously. A natural approach is weighted least square, minimizing sum of the squared distance  $\sum_i \omega_i (\theta-\theta)^2$ over $\theta$. However, since the angles are periodic, standard least square treats $\theta=0$ and $\theta=2\pi$ as different values even though they represent the same direction, which can lead to incorrect results. Therefore, we minimize the circular distance via the cost function,
\begin{eqnarray}
    {\cal L(\theta)} = \sum_i \omega_i \left ( 1- \cos(\theta-\theta_i)\right) 
    \label{eq:loss_wls}
\end{eqnarray}
where the weights $\omega_i$ are chosen as the sum of squared mode amplitudes in each bin
\begin{eqnarray}
    \omega_i = e_1(p_{T_i})^2+e_2(p_{T_i})^2 \ .
\end{eqnarray}
Minimizing ${\cal L(\theta)}$ is equivalent to maximizing $\omega_i \cos(\theta-\theta_i)$, which upon setting the derivative to zero yields the optimal global angle 
\begin{eqnarray}
    \theta = \tan^{-1} \left( \frac{\sum_i \sqrt{\omega_i} \ e_2(p_{T_i})}{\sum_i \sqrt{\omega_i} \ e_1(p_{T_i})} \right) \ .
\end{eqnarray}

\section{Alternative determination of $\theta$}
\label{s:rotation_alt}
Another independent approach to determine $\theta$ follows directly from the physical result established in the main text that $v_0(p_T)$ is driven purely by the thermal mode and the geometric mode carries negligible weight in $v_0(p_T)$.

From Eq.~(\ref{eq:reconst_v0v02_rotated}), the rotated weights are,
\begin{eqnarray} 
\tilde{w}^{v_0}_{T} &=& \cos\theta \ w^{v_0}_{1} + \sin\theta \ w^{v_0}_{2} \nonumber \\ 
\tilde{w}^{v_0}_{G} &=& -\sin\theta \ w^{v_0}_{1} + \cos\theta \ w^{v_0}_{2} \end{eqnarray} 

By imposing the condition $\tilde{w}^{v_0}_{G} \approx 0$ one directly obtains 
\begin{eqnarray} 
\theta = \tan^{-1}\left(\frac{w^{v_0}_{2}}{w^{v_0}_{1}}\right) \ .
\end{eqnarray} 
This is a single algebraic equation requiring no optimization, with $\theta$ determined entirely by the scalar PC weights $w^{v_0}_{1}$ and $w^{v_0}_{2}$ from Eq.~(\ref{eq:reconst_v0v02}). 

The two methods, maximal coherence and zero geometric weight in $v_0$, yield angles that agree to within 3 \% of the total rotation range ([0,$\pi$]), providing a strong self-consistency check of the framework.

\section{Experimental method for extractions of the modes}
\label{s:expt_method}
Here we describe a possible way in the experiment for the direct extraction of the physical modes described in the text.  To achieve this, an experiment needs to have four groups of sub-events recorded by four detectors sufficiently separated in pseudrapidity $\eta$ as shown in Fig.~\ref{fig:etabins}. This is similar to the method described in Appendix A. of the reference~\cite{Parida:2025eqv}. Let us label these four detectors from backward to forward rapidity as A, B, C, D.  
\begin{figure}[h!]
\includegraphics[width=0.5\textwidth]{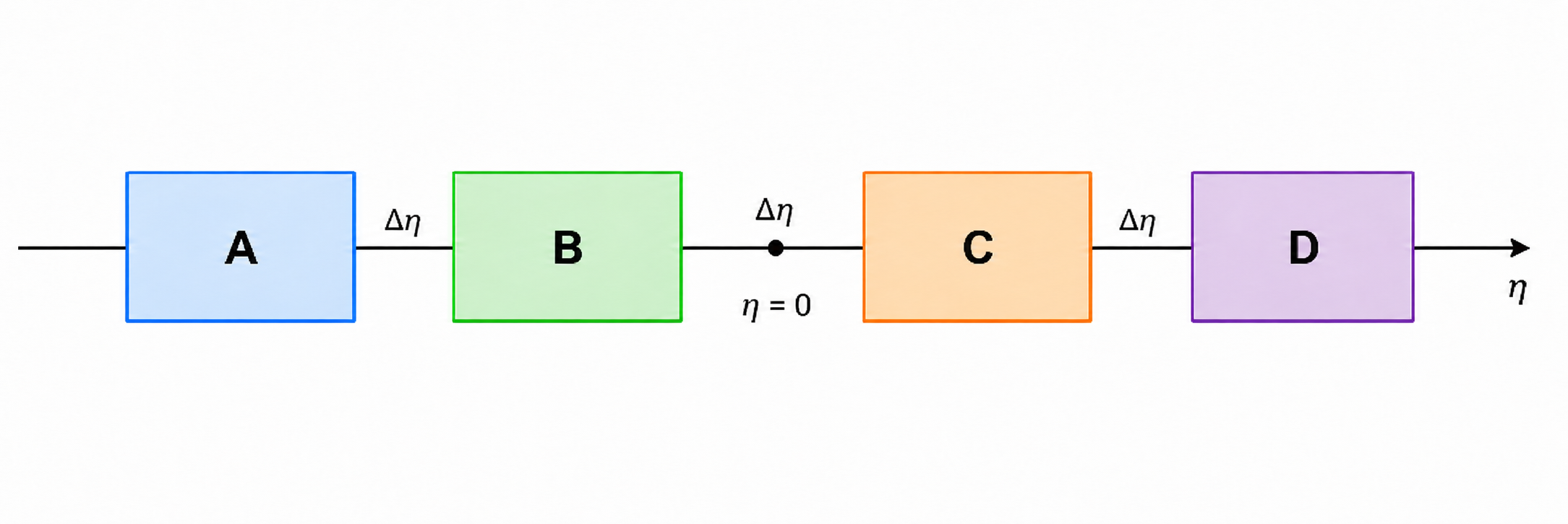} 
\caption{Schematics of the four pseudorapidity windows well-separated by $\delta \eta$ gaps to record events for measurement of observables of interest.~\label{fig:etabins}}
\end{figure}
Next, one needs to measure the core observables $N(p_T)$, $[p_T]$, $\sigma_{p_T}$, $v_2^2$ to be able to construct the blocks of the data covariance matrix similar to $\Sigma$.  

For event-by-event $v_2^2$, one needs to measure the Q-vectors~\cite{Bilandzic:2010jr} in the detectors A and D,
\begin{eqnarray}
    Q_{2,A} = \sum_{i=1}^{N_A} exp(2i \phi_i)\ , \\
    Q_{2,D} = \sum_{i=1}^{N_D} exp(2i \phi_i)\
\end{eqnarray}
where $N_A$ and $N_D$ are the number of particles in two detectors. Then for the normalized spectrum one measures the fraction of particles in a given $p_T$ bin in one of the other detectors, say B,
\begin{eqnarray}
    N(p_T)_B\equiv f_B(p_T) = \frac{N_B(p_T)}{N_B}.
\end{eqnarray}
Similarly, the detector C can be used to measure $f_C(p_T)$ and also event-by-event mean transverse momentum per particle $[p_T]$,
\begin{eqnarray}
    [p_T]_C = \frac{1}{N_C} \sum_{j=1}^{N_C} p_{T,j}.
\end{eqnarray}
Since we also need $\sigma_{p_T}$ for normalization of the covariance blocks, the best approach would be to measure mean transverse momentum per particle in another detector (say, $[p_T]_B$), which would suppress non-flow contamination~\cite{Parida:2024ckk},
\begin{eqnarray}
    \sigma_{p_T}^2 = \langle [p_T]_B [p_T]_C\rangle -  \langle [p_T]_B \rangle \langle  [p_T]_C\rangle \ ,
\end{eqnarray}
where $\langle \dots \rangle$ denotes the average over all events.

Then one constructs each block of the normalized data covariance matrix as given in Eq.~(\ref{eq:mat_elem}) as, 
\begin{eqnarray}
    C_{N,N} = \frac{\langle f_B(p_T) f_C(p_T)\rangle_c}{\langle f_B(p_T) \rangle \langle f_C(p_T)  \rangle} \ , \\
    C_{N,[p_T]} = \frac{\langle f_B(p_T) [p_T]_C\rangle_c}{\langle f_B(p_T) \rangle \sigma_{p_T}} \ , \\
    C_{N, v_2^2} = \frac{\langle f_B(p_T) Q_{2,A} Q_{2,D}^*\rangle_c}{\langle f_B(p_T) \rangle \langle Q_{2,A} Q_{2,D}^* \rangle_c} 
\end{eqnarray}
where the numerators are two- or three-particle cumulants with the subscript $c$ representing the corresponding connected parts of the correlations, after subtracting the uncorrelated or disconnected parts~\cite{Parida:2025eqv}.
It should be noted that measuring $C_{N,[p_T]}$ and $C_{N, v_2^2}$ is equivalent to measure $v_0(p_T)$ and $v_{02}(p_T)$. Therefore, only additional task would be to mesure $C_{N,N}$. Once each element or block of the normalized data-covariance matrix is obtained one can apply directly the PCA on that matrix similarly as has been done by CMS collaboration in ~\cite{CMS:2017mzx}. This should result in two leading eigenvectors explaining most of the fluctuations of spectra. Then, to identify the physical modes, one can use the fact established in this article that $v_0(p_T)$ is purely driven by thermal mode, to find the rotation angle $\theta$ following the method described in Appendix~\ref{s:rotation_alt}. 

The biggest challenge to perform the above analysis in experiment would be to design four sub-events in four well-separated $\eta$-windows. However, in case of overlapping pseudorapidity windows, one can remove the self-correlations systematically following the proceedure described at the end of refernce~\cite{Parida:2025eqv} and detailed in~\cite{DiFrancesco:2016srj}.

\end{document}